\documentclass[twocolumn,prl,superscriptaddress,floats,nobibnotes,notitlepage,citeautoscript]{revtex4-1}

\usepackage[pdftex]{graphicx}
\usepackage[utf8]{inputenc}
\usepackage{amsmath,bm}
\usepackage{bbold}
\usepackage{color,soul}
\usepackage{xstring}
\usepackage{wasysym}
\usepackage{xspace}
\usepackage{amssymb}
\usepackage{time}
\usepackage{bbold}
\usepackage{subfigure}
\usepackage{multirow}
\usepackage{xspace}
\usepackage{url}
\usepackage{endnotes}
\newcommand{\ve}[1]{\boldsymbol{#1}}

\usepackage[dvipsnames]{xcolor}
\definecolor{Zcolour}{rgb}{0.992, 0.588, 0.22}
\definecolor{Rcolour}{rgb}{0.60, 0.0, 0.67}

\setcitestyle{super}



\def\Tr{\mathop{\mathrm{Tr}}}

\makeindex

\begin{document}

\title{Phases of the (2+1) dimensional SO(5) non-linear sigma model with  topological term}

\author{Zhenjiu Wang }
\email{Zhenjiu.Wang@physik.uni-wuerzburg.de}
\affiliation{Institut f\"ur Theoretische Physik und Astrophysik, Universit\"at W\"urzburg, 97074 W\"urzburg, Germany}
\author{Michael P. Zaletel}
\affiliation{Department of Physics, University of California, Berkeley, CA 94720, USA}
\author{Roger S. K. Mong}
\affiliation{Department of Physics and Astronomy, University of Pittsburgh, Pittsburgh, PA 15260, USA}
\author{Fakher F. Assaad}
\email{assaad@physik.uni-wuerzburg.de}
\affiliation{Institut f\"ur Theoretische Physik und Astrophysik, Universit\"at W\"urzburg, 97074 W\"urzburg, Germany}
\affiliation{W\"urzburg-Dresden Cluster of Excellence ct.qmat, Am Hubland, D-97074 W\"urzburg, Germany} 

\begin{abstract}
We use the half-filled zeroth  Landau level  in graphene  as a regularization scheme to study the physics of the SO(5) non-linear sigma model   subject to a  Wess-Zumino-Witten topological term in 
2+1  dimensions.     As shown  by Ippoliti et al. [PRB 98, 235108 (2019)],   this approach  allows for  negative sign free  auxiliary field quantum Monte Carlo simulations. 
The model has  a single free parameter, $U_0$,  that  monitors  the stiffness.   Within the parameter range accessible to negative sign free simulations, we  observe an 
ordered phase   in the large $U_0$ or  stiff limit.    Remarkably, upon  reducing  $U_0$ the magnetization drops substantially, and the correlation length exceeds our biggest system sizes, accommodating 100  flux quanta.   
The  implications of  our results  for deconfined quantum phase transitions between   valence bond solids and anti-ferromagnets  are  discussed. 
\end{abstract}

\maketitle

\noindent
{\it Introduction.}  At critical points  the  renormalization group allows  for the definition of emergent symmetries and field theories.    
 For example,  the semi-metal to insulator  transitions in graphene \cite{Assaad13,Toldin14,Otsuka16}  have an emergent Lorentz symmetry 
\cite{Gross74,Herbut06}   so that space and time can interchangeably be used  \cite{Chandrasekharan13}  to efficiently compute  critical exponents.
Models that capture the physics of deconfined quantum criticality   (DQC) \cite{Senthil04_2,Senthil04_1} -- the JQ model for example \cite{Sandvik07}  --  have  an SO(3) $\times $ C$_4$ symmetry, but at criticality the  C$_4$  point group   is enlarged  to  a higher U(1) symmetry.
Improved models,   with SO(3) $\times$ U(1) symmetry  have been proposed to study DQC  \cite{Liu18}. 
Formulating  the theory of  DQC    with  eight component Dirac fermions
akin to graphene  and Yukawa coupled to a quintuplet of anti-commuting mass terms \cite{Ryu09}  quite naturally leads to the conjecture  of an emergent SO(5) symmetry  \cite{Abanov00,Tanaka05,Senthil06}.
Compelling numerical evidence for  this emergent symmetry  has been put forward \cite{Nahum15_1} in the realm of loop models \cite{Nahum15}.

   \begin{figure}
\centering
\includegraphics[width=0.5\textwidth]{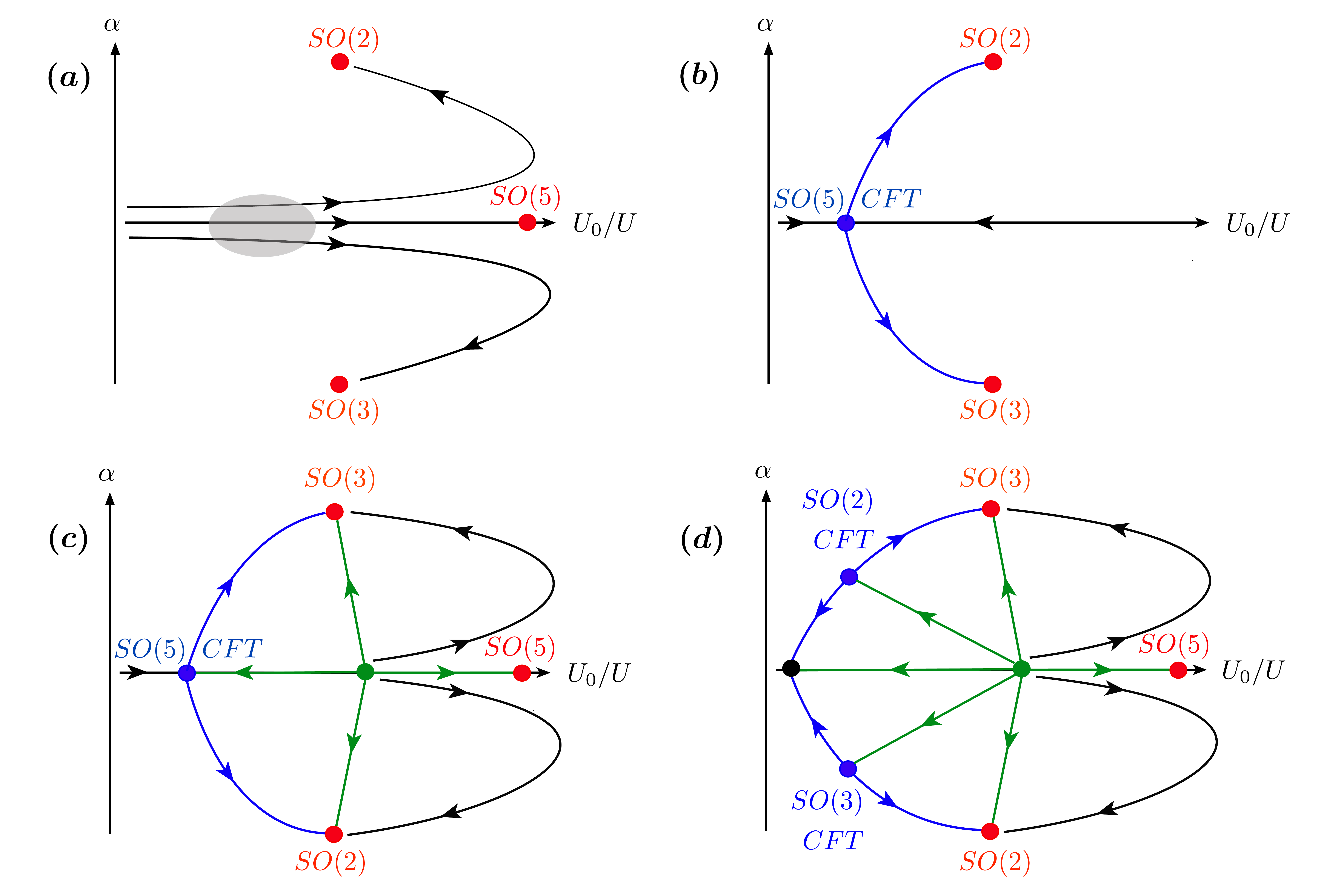} 
\caption{ 
 Possible RG flows in the  $U_0$ versus $\alpha$ phase.   $\alpha$ corresponds  to  the amplitude  of a term  that  breaks down the SO(5) symmetry to SO(3) $\times$ SO(2). The horizontal line corresponds to  the $\alpha =0$ or to SO(5) symmetry. 
  Red bullets, corresponds  to  phases where the  symmetry  group is spontaneously broken.   The black  bullet is an  SO(5)  disordered phase.  Blue (green)  bullets  denote  critical (multi-critical)  points.  In scenario (a)   the SO(5) model orders and the shaded region   depicts a slow RG flow (see text).  In (b)  the SO(5) model  remains critical.  In (c) the SO(5) model has an ordered and  critical phase separated by a multi-critical point. Finally, in (d) the SO(5) model shows an order-disorder transition. }
 \label{fig:RG_Flows}
\end{figure}

Let us now consider  a phase transition with enhanced symmetry and  a relevant  operator $\alpha$ which breaks it.
In this case,   formulating  a model with higher symmetry allows us to assess  the nature of the transition.
Schematic RG flows for an enhanced SO(5) symmetry  that is broken down to SO(3) $\times$ SO(2)  are shown in Fig.~\ref{fig:RG_Flows}.
While $\alpha$  breaks the SO(5) symmetry, $U_0$,   conserves it. 
If the higher symmetry model is in an ordered phase, then the transition is first order [Fig.~\ref{fig:RG_Flows}(a)].
The spin-flop transition   corresponding to the field-driven reorientation of the easy  axis falls into this category. 
Alternatively, the enhanced symmetry model  can be critical  such that the  transition is continuous [Fig.~\ref{fig:RG_Flows}(b)]. 
As an example for this scenario, we can consider the one-dimensional DQC   between  a dimer and N\'eel state  in the XXZ model considered in  Refs.~\cite{Haldane82,Mudry19} and realized in \cite{Weber19}.
The critical point has a U(1) symmetry that is broken by the  umklapp  operator that tunes through  the transition. 
As a third possibility,  the enhanced-symmetry model may have a relevant  tuning parameter   $U_0$ --  and associated (multi) critical point --   that does not break the  enhanced symmetry. 
Fig.~\ref{fig:RG_Flows}(c)  describes a scenario  where the ordered state gives way to a critical   phase. 
 In this case,  tuning $\alpha $  leads to first order or  continuous transitions depending upon the value of  $U_0$.
 Finally, in  Fig.~\ref{fig:RG_Flows}(d)    $U_0$  drives an order-disorder transition. 
 Aside from fine tuning, the transition from the  SO(3) to SO(2) broken symmetry  states   is first order or   separated by a disordered  phase.

The aim of this letter is to investigate the O(5)    non-linear sigma model in  2+1 dimensions  with a Wess-Zumino-Witten   geometrical term.    As mentioned above, this model can be obtained by 
integrating out   quintuplets of anti-commuting mass terms  in Dirac systems    \cite{Abanov00,Tanaka05,Senthil06}.  
\begin{equation}
\label{Eq:Action-Sigma}
    S = \frac{1}{g} \int d^3 x (\nabla \hat{\ve{\varphi}})^2     +   S_\text{WZW} .
\end{equation}
Here $\hat{\ve{\varphi}}$ corresponds to a five-dimensional unit vector.   The model has a manifest SO(5) symmetry, and a    single parameter, the stiffness. 
 The question we would like to  address in this work  is  the nature of the phase diagram as a function of the stiffness.   

{\it Model and method.}   The work of  Ippoliti et al. \cite{Ippoliti18}   demonstrates that a nonlinear sigma model with exact SO(5) symmetry can be constructed using $8$ component   Dirac fermions quenched in the zeroth Dirac  Landau level (ZLL). 
As opposed to a lattice  approach, one remains in continuum space time,   but the single particle Hilbert space remains finite and counts  the  $4 N_{\phi}$ states of the ZLL, where $N_{\phi}$  is the number of magnetic fluxes   piercing  the two dimensional space.     
 The model reads:
\begin{align} \begin{aligned}
 \hat{H} &=  \int_{V} d^2 \bm{x} \Bigg(  \frac{U_0}{2}  [\hat{ \ve{\psi}}^{\dagger} (\bm{x})  \hat{\ve{\psi}}(\bm{x}) - C(\bm{x} ) ]^2
 \\&\mkern90mu  -  \frac{U}{2}  \sum_{i=1}^{5} [ \hat{\ve{ \psi}}^{\dagger} (\bm{x}) O^i \hat{\ve{\psi } } (\bm{x})  ]^2  \Bigg)
 \label{Eq:Ham_r}
\end{aligned} \end{align}   
where the fermion annihilation operator are projected onto the ZLL:
$\hat{\psi}_a(\bm{x}) = \sum_{k=1}^{N_{\phi}}  \phi_k(\bm{x}) \hat{c}_{a,k}$.
The index $a$  runs from  $1 \cdots 4$ corresponding to the four Dirac ZLLs. 
It's crucial that the operators $\hat{\ve{\psi}}(\bm{x})$  ($\hat{\ve{\psi}}^{\dagger}(\bm{x})$) do not 
satisfy the canonical commutation rules due to the projection. 
The wave functions of  the  ZLL, $\phi_{k_y} (\ve{x})$, are computed in the  Landau gauge which diagonalizes the translation invariance along one  direction (see SM). 
The background $C(\bm{x}) \equiv 2\sum_{k_y=1}^{N_{\phi}} | \phi_{k_y}(\bm{x}) |^2 $ ensures particle-hole symmetry in Eq.~\eqref{Eq:Ham_r}.

For  $i=1 \cdots 5$,   $O^i$,   are  mutually  anti-commuting  matrices. 
 A convenient choice reads:  
\begin{equation}
 \tau_x \otimes  \mathbb{I}_2, \tau_y \otimes \mathbb{I}_2,  \tau_z \otimes \vec{\tau}  \ \ \   i=1,2,...5.  
 \label{Eq:five-matrix}
\end{equation}
The 10  matrices $L^{ij} = -\frac{i}{2}\left[ O^i, O^j   \right]$,  $i,j = 1 \cdots 5 $,  are the generators of  the SO(5) group
and  commute with the Hamiltonian.   Along the SO(5) high symmetry line,  the Hamiltonian has only one energy scale   $U_0/U$. 

Let  $ \varphi_{i}(\ve{x})  =  \langle  \hat{\ve{\psi}}^{\dagger} (\bm{x}) O^i \hat{\ve{\psi}}(\bm{x})  \rangle $ and assume that the mass gap  $\Delta  \propto | \ve{\varphi} | $ is finite.  One can then 
omit  amplitude fluctuations  of the vector $  \ve{\varphi}(\ve{x}) $,  integrate of the fermions  in the large mass approximation to obtain an effective field theory for $  \hat{ \ve{\varphi}}(\ve{x})   \equiv    \ve{\varphi}(\ve{x})/ | \ve{\varphi}(\ve{x}) |  $
that corresponds precisely to  Eq.~\eqref{Eq:Action-Sigma} \cite{Sachdev15}.  Here, we identify  $U_0/U$   to  $1/g$.

\begin{figure}
\centering
\includegraphics[width=0.45\textwidth]{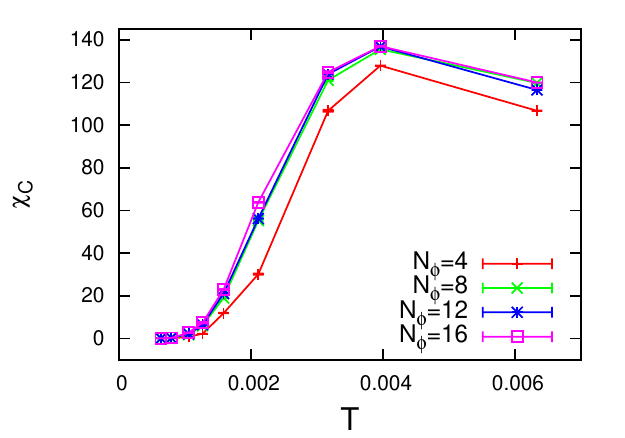}
\caption{ 
Temperature dependence of the uniform charge susceptibility $\chi_C$ for $U_0 = -1$.
}
\label{fig:Charge_Sus}      
\end{figure}

As mentioned previously,  our numerical simulations are based on the work of Ippoliti et al. \cite{Ippoliti18} that show how to formulate a negative  sign free auxiliary field QMC for  
the Hamiltonian of Eq.~\eqref{Eq:Ham_r} in the parameter range $U_0/U \geq -1 $.   The algorithm is formulated in Fourier space,  
\begin{equation} 
\hat{H}  = \frac{1}{2V}  \sum_{\ve{q}}  \left(   U_0 \hat{n}(\ve{q})  \hat{n}(-\ve{q}) -  U \sum_{i=1}^{5} \hat{n}^i(\ve{q})  \hat{n}^i(-\ve{q}) \right),
\end{equation}
with 
$  \hat{\ve{ \psi}}^{\dagger} (\bm{x}) O^i \hat{\ve{\psi }} (\bm{x}) = \frac{1}{V} \sum_{\ve{q}} e^{-i \ve{q} \cdot \ve{x} } \hat{n}^i(\ve{q})   $ and 
 $  \hat{\ve{ \psi}}^{\dagger} (\bm{x}) \hat{\ve{\psi }}(\bm{x}) - C(\bm{x}) = \frac{1}{V} \sum_{\ve{q}} e^{-i \ve{q} \cdot \ve{x} } \hat{n}(\ve{q})  $.  
 As shown in the supplemental material  (SM) one key point is to use a Fierz identity to avoid the negative sign problem.
 For each $\ve{q}$ we then  use a complex 
 Hubbard-Stratonovich  transformation to decouple the interaction term.  Auxiliary field QMC simulations   turn out to be  involved.    
The difficulty lies in the fact that the projected  density operators are not local  and that  they do not commute with each other.  In the SM we show that for a symmetric Trotter  decomposition,  that preserves the 
hermiticity  of the imaginary time propagation, the systematic error scales as $ \left( \Delta \tau \, N_\phi \right)^2 $.  Here we have set the magnetic unit length, $l_B$,  to unity such that 
$N_\phi$ corresponds to the volume.     We also note that the SO(5) symmetry is broken by the Trotter decomposition  such that it potentially introduces a relevant  operator. For all these reasons, great care has to be taken to control the  systematic error,  and we are obliged to scale $ \Delta \tau$ as  $1/N_\phi$.       A detailed account of the Trotter error is given in the SM.
 Since we are working in the continuum, the sum over momenta is not bounded. However,   the density operator  contains a factor  $e^{-\frac{1}{4}\ve{q}^2 l_B^2 }$  and  momenta  the exceed a critical value can be safely omitted.  Adopting this  regularization  strategy restricts the sum over momenta to  order $N_\phi$ values  again for the case $l_B=1$. Again a detailed  test of the choice of the momenta cutoff is given in the supplemental material.  
Taking all the  above into account yields a computational effort  that scales as  $N_\phi^5  \beta $ where 
$\beta$ is the inverse temperature.    This should be compared to the generic   $N_\phi^3  \beta $   scaling for say the Hubbard model.    The above explains why our simulations are limited to $N_\phi = 100$.
We have used the finite temperature auxiliary field 
algorithm \cite{Blankenbecler81,White89,Assaad08_rev}  of the  algorithms for lattice fermions (ALF)-library \cite{ALF_v1}.   The details of our implementation are discussed in the SM.

\textit{  Numerical results.}    For the simulations we  set the energy scale by choosing $U=1$,  the length scale by choosing  $ l_B = 1 $ and vary $U_0$ and the volume $N_\phi$.
We found that an inverse temperature of $\beta = 160 \pi^2 $   suffices to obtain ground state properties on our largest system sizes,  $N_\phi = 100$.  

In Fig.~\ref{fig:Charge_Sus}  we plot the uniform charge susceptibility, 
\begin{equation}
  \chi_C = \frac{ \beta}{N_{\phi}}  ( \langle \hat{n}_{\bm{q}=0}   \hat{n}_{\bm{q}=0}   \rangle 
   - \langle \hat{n}_{\bm{q}=0}  \rangle  \langle  \hat{n}_{\bm{q}=0}  \rangle ).
\end{equation}     

The charge fluctuations decay exponentially upon reducing the temperature  as expected for an  insulating state of matter. 
Since $\hat{\ve{ \psi}}^{\dagger} (\bm{x}) O^i \hat{\ve{\psi }} (\bm{x}) $ are mass terms,   any non-vanishing expectation  value of these fermion  bi-linears, $\varphi_i$,  will lead to a charge gap.    Owing to the SO(5) symmetry the single  particle gap is proportional to the norm of this  five component vector, $| \ve{\varphi} | $. 

\begin{figure}
\includegraphics[width=0.45\textwidth]{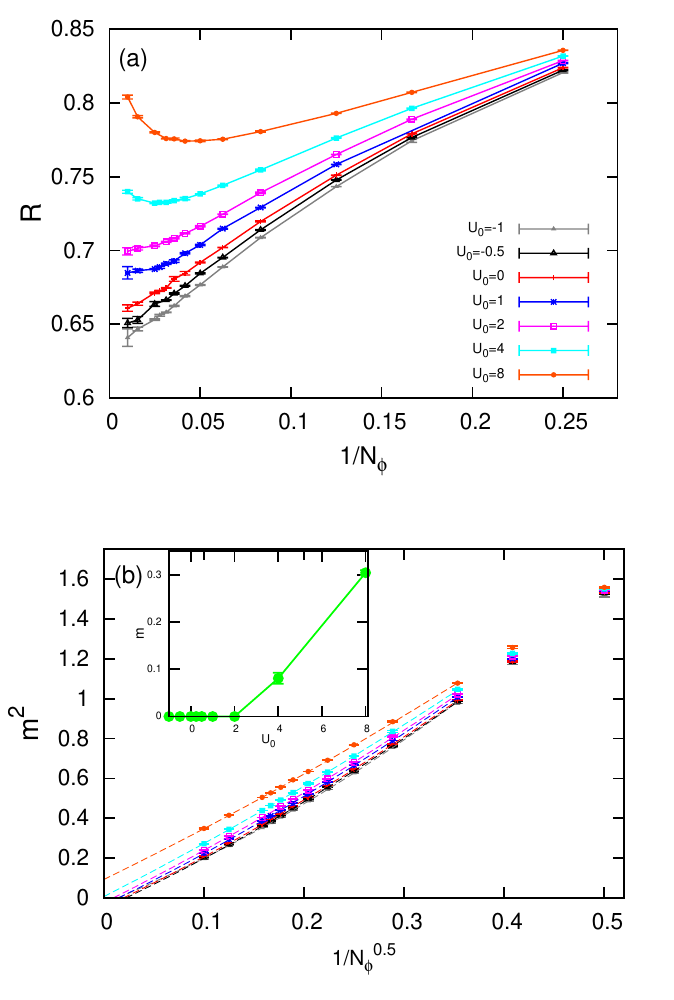} \\
\includegraphics[width=0.45\textwidth]{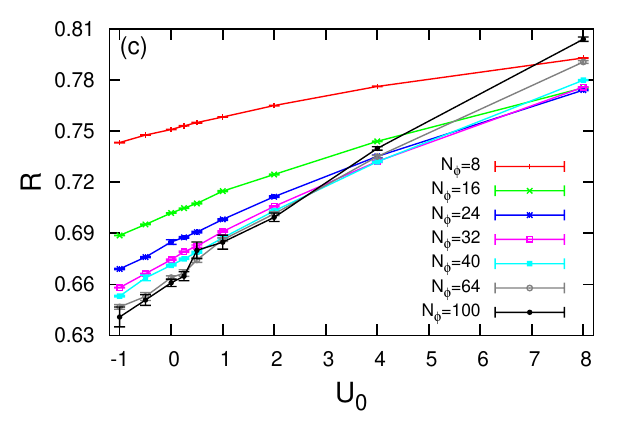} 
\vspace{-3ex}
\caption{ 
Correlation ratio $R$ as a function of $1/N_{\phi}$ (a) and $U_0$ (c), as well as the O(5) order parameter (b) as a function of $1/\sqrt{N_{\phi}}$.
The dashed lines and the inset is the extrapolation via the fitting form $m(N_\phi)^2 = m^2_{N_{\phi} \rightarrow \infty} + a / \sqrt{N_{\phi}} + b / N_{\phi}$. 
 Negative extrapolated values of $m^2$   suggest a critical or  disordered state. In both cases the  polynomial, in inverse linear length,  fitting form is not justified.
}
\label{fig:SO5_R_m}
\end{figure}

Although $| \ve{\varphi} | $ is  finite,  phase fluctuations  can destroy ordering. To numerically investigate this possibility,  we compute the  order parameter correlation function
\begin{equation}
  S(\ve{q})  = \frac{1}{N_{\phi}}  \sum_{i=1}^5  \langle    \hat{n}^i_{\ve{q}} \hat{n}^i_{-\ve{q}}     \rangle.
\end{equation}
For an ordering wave vector $\ve{Q}$,   the local moment reads 
\begin{equation}
   m = \sqrt{\frac{1}{N_{\phi}} S(\bm{Q})  }    
\end{equation} 
and it is convenient to define a renormalization group invariant quantity
\begin{equation}
    R \equiv 1 - \frac{S(\bm{Q} + \Delta \bm{q}) }{ S(\bm{Q}) }
\end{equation}
with $  | \Delta \ve{q} | = \frac{2 \pi}{\sqrt{N_\phi}} $.
In  the ordered (disordered)  phase $R$ converges to unity (zero)  and the  local moment takes a finite (vanishing) value.   At a critical point, 
the correlation ratio  converges to a universal   value. 

In Fig.~\ref{fig:SO5_R_m}(a) we plot the correlation ratio  $R$ as a function of system size  for various values of $U_0$.   For system sizes up to $N_\phi = 20 $ all curves scale downwards  and 
would suggest a  critical or disordered phase. Beyond $N_\phi = 20 $   and for  large values of $U_0$  the correlation ratio    changes behavior and grows.  The length scale 
at which this crossover  occurs  can naturally be interpreted as a measure of the correlation length. For   these large values of $U_0$, a finite size extrapolation of the square of the 
local moment  (see Fig.~\ref{fig:SO5_R_m}(b)) is consistent with a finite value   (see inset of Fig.~\ref{fig:SO5_R_m}(b)).
In  Fig.~\ref{fig:SO5_R_m}(c)  we replot the  correlation ratio as a function of $U_0$.    The data is consistent with a crossing at $U_0  \simeq 3$. Below this value,   $R$ does not scale to zero, as already seen in Fig.~\ref{fig:SO5_R_m}(a),  and hence signals a phase where the correlation length exceeds our system sizes.

\textit{ Discussion. }   In Fig.~\ref{fig:RG_Flows}    we show possible RG flows in the $U_0$ versus  $\alpha$ plane where $\alpha$ corresponds  to  the amplitude  of a term  that  breaks down the SO(5) symmetry to SO(3) $\times$ SO(2).   

Fig.~\ref{fig:RG_Flows}(a) corresponds a scenario where  the  topological term is  irrelevant and the model orders for all values of the stiffness.     Taken at face value, our results do not support this  point of view.  However we cannot exclude the possibility that  an ordered phase with small magnetic moment  will occur on larger system sizes.   In this case, the transition as a function of $\alpha$   from the SO(3) to SO(2)  broken symmetries   corresponds to a spin-flop transition.

In contrast  in Fig.~\ref{fig:RG_Flows}(b)  we assume  that the SO(5) model corresponds to a CFT. In this case,  $\alpha$  is a relevant parameter, and  the transition from SO(3) to SO(2)  broken symmetry phases  is continuous with an emergent SO(5) CFT  at the critical point.  This SO(5) CFT could be a candidate theory for DQCP.  Again  in light of our data, this scenario seems unlikely since at large values of  $U_0$ our data supports an ordered phase.

In Fig.~\ref{fig:RG_Flows}(c)  we assume that the observed ordered phase gives way to  a critical phase corresponding to an  SO(5) CFT.   Adding the $\alpha$  axis implies that  along the SO(5) line  we have a multi-critical point as well as  a  critical one. Our data actually favors this scenario:  below $ U_0  = U_0^{c}   \simeq 3 $  the correlation  ratio does not  seem to scale to zero, and is hence consistent  with a critical phase.  If such is the case,  the nature of the  transition between SO(3) and SO(2)  broken symmetry states,  with emergent SO(5) symmetry,  depends upon the value of $U_0$  and is either continuous or first order.   
There are a number of models that show a transition from  SO(2) (VBS/SSC)  to SO(3) (AFM/QSH) broken symmetry phases and that favor continuous or weakly first order quantum phase  transitions.  
For instance, 3D loop models \cite{Nahum15}
($\eta_\text{Neel} = 0.259(6)$, $\eta_\text{VBS} = 0.25(3) $),  the J-Q model, \cite{Sandvik07,Shao15} as well as transitions between  quantum spin Hall insulators and $s$-wave superconductors, 
 ($\eta_\text{QSH}=0.21(5)$, $\eta_\text{SSC}=0.22(6)$)  \cite{Liu18} all seem to show similar exponents and are believed to belong to  the class  of  DQCP  with emergent   spinons coupled to a non-compact U(1) gauge field.  Compelling evidence of emergent SO(5) symmetry has been put  forward for the loop model \cite{Nahum15_1}. 
  However,   the value of  the anomalous dimension lies at odds  with conformal bootstrap bounds, $ \eta > 0.52$ \cite{Poland18}, for emergent SO(5) symmetry.   Systematic drift in the  exponents  has been observed in \cite{Nahum15}.  Within the present context one can understand  the above in terms fix-point collision  put forward in \cite{WangC17,Nahum19,WangC19,Janssen14}.
Consider a third axis --  the dimension -- and assume that  the sketch of Fig.~\ref{fig:RG_Flows}(c)   is realized \emph{close}  to the physical dimension $d=2$ but that  before  approaching $d=2$ the multi-critical and critical points collide and develop a complex component.   In this case we are back to the spin-flop transition of  Fig.~\ref{fig:RG_Flows}(a)    but  with the important insight   that the RG flow becomes arbitrarily slow   due to  proximity  of  a fix-point collision.   The shaded 
  region in  Fig.~\ref{fig:RG_Flows}(a)  schematically depicts the region where the RG flow becomes very slow.  
    
Proximity to a critical point motivates fitting  the QMC data to the form: $ m = m_0 + a N_{\phi}^{-\frac{\eta +z}{4} }  $.   In the region where  our correlation length exceeds the size of our system we obtain a good fit with robust anomalous dimension $\eta = 0.28(2) $  under the  assumption of $z=1$  (see the SM).    The  agreement with the aforementioned QMC results  is remarkable.    We note that this exponent is much larger than the one of the   $3D$ classical O(5)
    critical point, with  $\eta=0.036(6)$ \cite{LIU2012107}. 
 We  conclude this section  by noting that  Ref.~\onlinecite{Yao19} introduces a fermion model showing a DQCP  with  emergent SO(5) symmetry  and that has exponents that comply  with the bootstrap bounds.  This model could be a realization of the SO(5) CFT conjectured in Fig.~\ref{fig:RG_Flows}(c).

 Fig.~\ref{fig:RG_Flows}(d)     describes the possibility of an order-disorder transition  along the SO(5) line.  Note however that  on  the accessible system sizes, we cannot resolve the  length scale  associated  with the disordered state. This scenario excludes a DQCP with emergent SO(5) symmetry, and  the transition from the disordered to ordered phases involve SO(3) or SO(2)  critical points.  
 As shown in Fig.~\ref{fig:Charge_Sus}  the insulating phase has  vanishing charge susceptibility. 
 The existence  and nature of an SO(5)  symmetric  disordered phase is  intriguing. 
 Starting  from Dirac fermions  any band insulating state necessarily involves   SO(5) symmetry breaking.
 Hence in the conjectured phase diagram of Fig.~\ref{fig:RG_Flows}(d)    the disordered phase   is not adiabatically connected to a  band insulator. 
  In fact, if the disordered phase preserves the  particle-hole symmetry, the arguments of Ref.~\onlinecite{WangC17} rule out \emph{any} gapped phase (even a topological one), because the particle-hole symmetry forbids the SO(5) Hall-conductance argued to be necessary in any such insulator.

\textit{Conclusions -- }   Our data on systems up to $N_\Phi = 100$  show  that the SO(5) non-linear sigma model  exhibits an ordered phase in the limit of  large stiffness.  Remarkably (and within the  accessible  parameter range where negative sign free AFQMC simulations can be carried out), we observe another  regime  characterized by  a  correlation length that exceeds our system size. 
Given the aforementioned  body of work on DQC and insights from the  conformal bootstrap approach, our results find a natural interpretation by assuming that the  model lies close to a  fix-point with small complex component \cite{WangC17,Nahum19,WangC19}  such that the RG flow becomes very slow and shows pseudo-critical behavior. 
Clearly larger system sizes are desirable  so as to confirm this point of view. 
Although very appealing,  as implemented the Landau level projection approach   comes with a computational effort that scales as  $N_\phi^5 \beta$ as opposed to $N_\phi^3 \beta$  for the generic Hubbard model. 
Further improvements to the code will have to be  implemented  so as to reach  bigger  flux values.
The method can also be applied to the O(4) model with   $\theta$-term at $\theta=\pi$  by setting one mass term to zero.
This will have impact on  our understanding of easy plane de-confined quantum critical points with emergent O(4) symmetry. 

{\begin{acknowledgments}
We would like to thank A. Nahum, W. Guo, T. Senthil and C. Wang for discussions, as well as Matteo Ippoliti  with whom we have carried out previous work on this subject.  
The authors gratefully acknowledge the Gauss Centre for Supercomputing e.V. (www.gauss-centre.eu) for funding this project by providing computing time on the GCS Supercomputer SUPERMUC-NG at Leibniz Supercomputing Centre (www.lrz.de). 
FFA  thanks funding from the Deutsche Forschungsgemeinschaft    under the grant number AS 120/15-1  as well as through the W\"urzburg-Dresden Cluster of Excellence on Complexity and Topology in Quantum Matter - ct.qmat (EXC 2147, project-id 39085490).
 ZW thanks financial support from the DFG funded SFB 1170 on Topological and Correlated Electronics at Surfaces and Interfaces.  
MZ was supported by the DOE, office of Basic Energy Sciences under contract no.\ DEAC02-05-CH11231.
RM was supported by the National Science Foundation under grant No.\ NSF-1848336.
\end{acknowledgments}}

\bibliography{fassaad}

\clearpage

\section{Supplemental material }

\maketitle

\subsection{ Landau Level projection}

We consider  electrons  confined to the two-dimensional x-y plane  and in a  transverse magnetic field: 
\begin{equation}
  \hat{H}_0   = \frac{1}{2m}   \left(  \hat{\ve{P}} - e \ve{A}(\ve{r})\right)^2. 
\end{equation}
In  the  Landau gauge,  $\ve{A}(\ve{r})   = B(0,x,0)$   ($\ve{r} = (x,y,z) $),   
translations along the $y$ direction leave the Hamiltonian invariant  such that  the momentum in this axis,  $p_y$,  is a good   quantum number.
On a torus of size $L_x  \times L_y$  the wave function of the first    Landau level reads:   
\begin{equation}
 \phi_{p_y} (\bm{r}) = \frac{1}{\sqrt{L_y}} \frac{1}{\sqrt{l_B \sqrt{\pi}}} 
     e^{-(x/l_B - \text{sign}(B)  p_y l_B)^2 /2}  e^{i p_y y}.
\end{equation}
Here the magnetic length scale is defined as $l_B^2 = \frac{ \phi_0 }{2 \pi | B|} $, with $\phi_0 = \frac{h}{e}  $ ,
and the number 
of magnetic fluxes piercing the system, $N_{\phi} = \frac{|B| V}{\phi_0} $,  is an  integer  so as to guarantee uniqueness 
of the wave function.    Finally   the momentum  in the y-direction  is given  $p_y = \frac{2 \pi n}{L_y} $  with $n \in 1, \cdots,  N_{\phi}$.     From here  onwards we will consider 

The orbital wave function of the first Landau level of free electrons in a magnetic field  and that of the  zero energy Landau level  (ZLL)  in graphene are identical. In graphene, however, there is an SU(4)  symmetry such the  electron carries an additional  flavor index, $a  \in 1, \cdots, 4 $. Let  $\hat{c}_{p_y,a}$    destroy an electron in the  ZLL  with  flavor index $a$ and momentum $p_y$.  These operators satisfy   canonical fermion  commutation rules:
\begin{equation}  
\left\{ \hat{c}^{\dagger}_{p_y,a},   \hat{c}^{}_{p'_y,a'} \right\} = \delta_{a,a'}\delta_{p_y,p'_y},    \, \,  \, 
\left\{ \hat{c}^{}_{p_y,a},   \hat{c}^{}_{p'_y,a'} \right\}  = 0.
\end{equation}

Our Hamiltonian is defined in terms of projected field operators. 
\begin{equation}
\hat{\psi}_a( \ve{r}) = \sum_{p_y=1}^{N_{\phi}} \phi_{p_y}(\bm{r})  \hat{c}_{a,p_y}.
\end{equation}  
Since the ZLL  does not span the Hilbert space  the projected field operators do not satisfy the fermion canonical commutation  rules, and before formulating the AFQMC we have to  express everything in terms of the  canonical operators  $\hat{c}_{a,p_y}$.     Defining the    Fourier transform of the four component spinor:
\begin{equation}
	\hat{\ve{\psi}}_{\ve{p}}^{\dagger}   = \frac{1}{\sqrt{V}} \int_V d^2\ve{r} e^{ i \ve{p} \cdot \ve{r}} 
	\hat{\ve{\psi}}^{\dagger}(\ve{r})
\end{equation}
we  obtain 
\begin{equation}\label{Eq:Ham_Fourier}
\begin{aligned}
  \hat{H} & = \sum_{i=0}^5 \int_V d^2 \bm{r} \frac{U_i}{2}  [ \hat{\psi}^{\dagger}_a (\bm{r}) O^i_{ab} \hat{\psi}_b (\bm{r}) 
   - C(\bm{r}) \delta_{i,0} ]^2      \\    
 & = \sum_{i=0}^5  \frac{U_i}{2}  \sum_{\bm{q}} \sum_{\bm{q'}}  \int_V d^2 \bm{r}  
   (\frac{1}{V} e^{i \bm{q} \cdot \bm{r} } \hat{N}^i(\bm{q}) )    (\frac{1}{V} e^{i \bm{q}' \cdot \bm{r} } \hat{N}^i(\bm{q}') )   \\ 
 & = \frac{1}{2V} \sum_{i=0}^5  \sum_{\bm{q}}   \hat{N}^i(\bm{q}) U_i  \hat{N}^i(-\bm{q})  
\end{aligned}
\end{equation}  
Here,  $O^{0}$, is the unit matrix  and  
$  \hat{\ve{ \psi}}^{\dagger} (\bm{r}) O^i \ve{\hat{\psi }} (\bm{r}) = \frac{1}{V} \sum_{\ve{q}} e^{-i \ve{q} \cdot \ve{r} } \hat{N}^i(\ve{q})   $  for  $i=1 \cdots 5 $ and 
 $  \ve{ \hat{\psi}}^{\dagger} (\bm{r}) \ve{\hat{\psi} } (\bm{r}) - C(\bm{r}) $ $  = \frac{1}{V} \sum_{\ve{q}} e^{-i \ve{q} \cdot \ve{r} } \hat{N}^0(\ve{q})  $

Neglecting the constant background term at $i=0$, the  \textit{density} operators $ \hat{N}^i(\bm{q})$,  can be expressed in terms of the  canonical operators  $  \hat{\ve{c}}^{\dagger}_{p_y}$: 
\begin{equation}
\begin{aligned} 
&\hat{N}^{i}(\ve{q} ) =  \sum_{\bm{p}}  \hat{\ve{\psi}}^{\dagger}_{\bm{p}}  \ve{O}^i \hat{\ve{\psi}}^{}_{\bm{p}- \bm{q}}   \\  
= & \frac{1}{V}  \sum_{\bm{p}} \int_V \int_{V'} 
  d^2 \bm{r}  d^2 \bm{r'}   e^{i \bm{p} \cdot \bm{r}}  e^{-i (\bm{p} - \bm{q})\cdot \bm{r'}}   \\ 
  & \sum_{k_1} \hat{\ve{c}}^{\dagger}_{k_1}  \phi^{*}_{k_1} (\bm{r})  \ve{O}^i  \sum_{k_2} \hat{\ve{c}}_{k_2} \phi_{k_2}  (\bm{r'})    \\   
= & \frac{1}{V}  \sum_{\bm{p}}  \sum_{k_1} \sum_{k_2} \hat{\ve{c}}^{\dagger}_{k_1} \ve{O}^i  \hat{\ve{c}}_{k_2} 
 \int_V \int_{V'}  d^2 {\bm{r}}  d^2 {\bm{r'}}  \\ 
 & \left( \frac{1}{\sqrt{L_y}} \frac{\pi^{-\frac{1}{4}} }{\sqrt{l_B}} e^{-i {k_1} y }  e^{ -\frac{1}{2} (\frac{x }{l_B} -k_1 l_B )^2 }  e^{i \bm{p} \cdot \bm{r}  } \right)   \\ 
\cdot  & \left( \frac{1}{\sqrt{L_y}} \frac{\pi^{-\frac{1}{4}} }{\sqrt{l_B}} e^{+i {k_2} y'}  e^{ -\frac{1}{2} (\frac{x'}{l_B} -k_2 l_B )^2 }  e^{i (\bm{p} - \bm{q} ) \cdot \bm{r'} } \right)    \\     
= & \sum_{\bm{p}}  \frac{2l_B {\pi}^{1/2}}{L_x}  e^{i p_x p_y l_B^2} e^{-l_B^2 p^2_x /2 }       \\ 
  & e^{i(p_x - q_x)(p_y -q_y)l_B^2 }  e^{-l_B^2 (p_x - q_x)^2 /2 }   \hat{\ve{c}}^{\dagger}_{p_y} \ve{O}^i \hat{\ve{c}}_{p_y - q_y}     \\  
= & \frac{1}{2 \sqrt{\pi}}  e^{-l_B^2 \bm{q}^2 /4}  \sum_{p_y} e^{i q_x p_y l_B^2 }  
  \hat{\ve{c}}^{\dagger}_{p_y +\frac{q_y}{2}}  \ve{O}^i  \hat{\ve{c}}_{p_y - \frac{q_y}{2}}
\end{aligned}    
\end{equation}
In the last step, the sum over $p_x$  is carried  out  by changing sums to integrals and taking the limit $L_x \rightarrow  \infty $. 
With  the substitution $k = p_y + \frac{q_y}{2} $ and  
\begin{equation}\label{Eq:density}   
\begin{aligned}
      \hat{n}^i (\bm q )    
   = &   \sum_{k=1}^{N_{\phi}} \sum_{a,b=1}^4 F(\bm q)  e^{ \frac{i}{2} (2 k - q_y) l_B^2 q_x }    
     ( \hat c^{\dagger}_{a,k} O^i_{a,b}  \hat c_{b, k - q_y}   \\ 
    & - 2\delta_{q_y,0}\delta_{i,0} ) 
\end{aligned}     
\end{equation}
the Hamiltonian reads:
\begin{equation}
	\hat{H}  = \frac{1 }{8 \pi V} \sum_{i=0}^5  \sum_{\bm{q}}   \hat{n}^i(\bm{q}) U_i  \hat{n}^i(-\bm{q}).  
\end{equation}
In the above, $a($b$) =1,2,3$ and $4$  is the flavor index,  and 
 $F( {\bf q} ) \equiv e^{-\frac{1}{4} (q^2_x + q^2_y)l^2_B  }$.   
 The background term $2 \delta_{q_y,0} \delta_{i,0} $  can  easily be  verified by Fourier transform 
 the real space background $C(\bm{r})$ (see main text).

 As we will shown in the next subsection, this exponential decaying factor is essential for the QMC simulation since 
 it provides a   natural cutoff for the momenta $\ve{q}$.      Finally,  setting the magnetic unit length to unity  such that  $\frac{2\pi}{V} = \frac{1}{N_{\phi}} $  we obtain:

\begin{equation}
 	\hat{H}  = \frac{1}{16 \pi^2 N_{\phi}} \sum_{i=0}^5  \sum_{\bm{q}}   \hat{n}^i(\bm{q}) U_i  \hat{n}^i(-\bm{q})  
\end{equation}


\subsection{ Fierz identity and absence of the negative sign problem } 
To  avoid the negative  sign problem  in the QMC simulations we use the   Fierz identity to rewrite Eq.~\eqref{Eq:Ham_Fourier} as:
\begin{equation}
\label{Eq:Fierz}
\begin{aligned}
   H  = \frac{1}{ 2 N_{\phi}} \sum_{i=0}^3  \sum_{\ve{q}}  \hat{n}^i_{-\bf q} g^i \hat{n}^i_{\bf q}  \\
\end{aligned}
\end{equation}
Instead of the original density operators in Eq.~\eqref{Eq:density}, the $n^i$($i=0,1,2,3$) operators are based on $4$ matrix:
\begin{equation}
   \mathbb{I}_4, \tau_x \otimes \mathbb{I}_2,  \tau_y \otimes \mathbb{I}_2, \tau_z \otimes \mathbb{I}_2.
\end{equation}
Eq.~\eqref{Eq:Fierz} is identical to Eq.~\eqref{Eq:Ham_Fourier} when $\frac{g^0}{8 \pi^2} = U_0 + U$, $\frac{g^1}{8 \pi^2} = \frac{g^2}{8\pi^2} = -2U $, and $\frac{g^3}{8\pi^2} = 2 U$.  
Here we consider the SO(5) symmetric point and set  $U_i = -U$  for $i \in 1,\cdots, 4 $.
The absence of sign problem holds for the region of $U_0 \geq -U$,  
follows from the work of Ref.~\cite{Li16} and is discussed in detail in reference \cite{Ippoliti18}.   
The above matrix structure also gives an explicit $SU(2)$ symmetry which holds for  each  
Hubbard-Stratonovich field  configuration.  

\subsection{Trotter errors}

Since $n(\bm q)^{\dagger} = n(- \bm q)$, the exponential of operators at each time slice is  given by:   
\begin{equation}
 \begin{aligned}
    & e^{ -\frac{\Delta \tau }{2 N_{\phi}} ( \hat{n}^i_{\bm q} g^i \hat{n}^i_{-\bm q} + \hat{n}^i_{-\bm q} g^i \hat{n}^i_{\bm q} ) }   \\ 
  = & e^{ -\frac{\Delta \tau }{4 N_{\phi}} [g^i(\hat{n}^i_{\bm q} + \hat{n}^i_{-\bm q} )^2  - g^i (\hat{n}^i_{\bm q} -\hat{n}^i_{-\bm q} )^2 ] }
 \end{aligned}  
\end{equation}

To ensure hemiticity,  we use a symmetric Trotter decomposition:
\begin{equation}
  Z = \Tr{ [ \prod_{m=1}^{N}  e^{-\frac{\Delta \tau}{2} \hat{H}_m} 
             \prod_{n=N}^{1}  e^{-\frac{\Delta \tau}{2} \hat{H}_n} ]^{L_{\tau}} }
\end{equation}
where $\hat{H}_m$  corresponds to the $N = 2 \times 4  \times  N_{q} $  operators
$ \pm \frac{g^i}{4 N_{\phi}} (\hat{n}^i_{\bm{q}} \pm \hat{n}^i_{-\bm{q}} )^2  $. 
$N_q$ is the number of momentum points used for the simulation.  As we will see below $N_{q}$ scales as $N_\phi$.   

For two operators $\hat{H}_1 $ and $\hat{H}_2 $   the  leading  order error produced in the symmetric Trotter decomposition reads:
\begin{equation}
\begin{aligned}
   & e^{ -\frac{\Delta \tau}{2} \hat{H}_1 }  e^{  -\Delta \tau \hat{H}_2 }  e^{ - \frac{\Delta \tau}{2} \hat{H}_1 }   \\   
 = & e^{ -\Delta \tau (\hat{H}_1 + \hat{H}_2) + \frac{\Delta \tau^3}{12} [ 2 \hat{H}_1 + \hat{H}_2, [\hat{H}_1, \hat{H}_2  ] ] }  +  \mathcal{O} (\Delta \tau^4)
\end{aligned}
\end{equation}

Iterating the above formula gives: 
\begin{equation} \label{Eq:iterated-trotter}
\begin{aligned}
    &   \prod_{m=1}^{N}  e^{-\frac{\Delta \tau}{2} \hat{H}_m}      \prod_{n=N}^{1}  e^{-\frac{\Delta \tau}{2} \hat{H}_n}    
  = &  e^{- \Delta \tau((\sum_{m=1}^N \hat{H}_m )  +  \hat{\lambda} ) }   
  + & \mathcal{O} (\Delta \tau^4 ) 
\end{aligned}
\end{equation}
where 
\begin{equation}
\label{Eq:Trotter_lambda}
\begin{aligned}
    \hat{\lambda}  & \equiv -\frac{ \Delta \tau^2 }{12} ( \sum_{m=1}^{N-1}  \sum_{m'=m+1}^N  [2 \hat{H}_m + \hat{H}_{m'}, [\hat{H}_m , \hat{H}_{m'}] ]    \\  
  & +  \sum_{m=1}^{N-1}  \sum_{m'=m+1}^N   \sum_{m'' =m +1 }^N  [ \hat{H}_{m'},  [ \hat{H}_m,  \hat{H}_{m''}] ]  (1 - \delta_{m', m''} )  )
\end{aligned}
\end{equation}
and $\delta_{m',m''}$  the Kronecker delta.  
Using time dependent perturbation theory, we  then obtain: 
\begin{equation}
\begin{aligned}
   & \left( \prod_{m=1}^{N}  e^{-\frac{\Delta \tau}{2} \hat{H}_m}      \prod_{n=N}^{1}  e^{-\frac{\Delta \tau}{2} \hat{H}_n}  \right)^{L_{\tau}}  \\ 
 = &  e^{- \beta \hat{H} } - e^{ -\beta \hat{H}} \int_0^{\beta}  d \tau  e^{\tau \hat{H}} \hat{\lambda}  e^{-\tau \hat{H}}   +  \mathcal{O} (\Delta \tau^3 )
\end{aligned}
\end{equation}
with $L_{\tau}=\frac{\beta}{\Delta \tau}$  the number of time slices. 
$\hat{\lambda}$   is measure of the   leading order error on the free energy density: 
\begin{equation}
\begin{aligned}
 f_{QMC} \equiv &  -\frac{1}{\beta V} \ln{ \Tr{  ( \prod_{m=1}^{N}  e^{-\frac{\Delta \tau}{2} \hat{H}_m}    
   \prod_{n=N}^{1}  e^{-\frac{\Delta \tau}{2} \hat{H}_n}  )^{L_{\tau}}  } }      \\  
 =&  f + \frac{1}{\beta V} \int_0^{\beta} d\tau  
   \langle  e^{\tau \hat{H}}  \hat{ \lambda}   e^{- \tau \hat{H}}  \rangle   + \mathcal{O} (\Delta \tau^3 )   \\  
 =&  f  +  \frac{1}{ V }  \langle  \hat{\lambda}  \rangle  +  \mathcal{O} ( \Delta \tau^3 )      \\
 =&  f  +  \frac{1}{ 2 \pi N_{\phi} }  \langle  \hat{\lambda}  \rangle  +  \mathcal{O} ( \Delta \tau^3 )  
\end{aligned}
\end{equation}
In the above we have set  $ l_B = 1$ so as to replace $V$ by $N_\phi$,  and  $ f   = - \frac{1}{\beta V }  \ln \Tr{e^{-\beta \hat{H}}}$.  Since the interacting operators  for different  masses  $i$  do not commute with each other, the Trotter decomposition  
breaks the SO(5) symmetry of Hamiltonian (a SU(2) symmetry is left due to the Fierz identity in Eq.~\eqref{Eq:Fierz}).

To evaluate the expectation value of $\hat{\lambda}$,  we first evaluate the commutator of two density operators: 
\begin{equation}
\begin{aligned}\label{Eq:density_commutation}
   & [\hat{n}^i(\bm{q_1}) , \hat{n}^j(\bm{q_2})]   \\
 = & F(\bm{q_1}) F(\bm{q_2})  \sum_k  \ve{\hat{c}}^{\dagger}_k   \{ 
     e^{\frac{i}{2} (2k - (q_{1y}+ q_{2y})) l^2_B (q_{1x} + q_{2x}) }    \\  
   & (2 \cos ( \theta_{\bm{q_1}, \bm{q_2}} )  [O_i, O_j] +   2 i \sin ( \theta_{\bm{q_1}, \bm{q_2}} )  \{O_i, O_j \} )   \}  
   \ve{\hat{c}}_{k-(q_{1y}+q_{2y})}  \\ 
 = & \frac{F(\bm{q_1}) F(\bm{q_2}) }{F(\bm{q_1} + \bm{q_2})}  
   \{ n^{[O_i, O_j]}    (\bm{q_1} +\bm{q_2})  2  \cos ( \theta_{\bm{q_1}, \bm{q_2}} )    \\
  + & 
      n^{\{O_i, O_j\} } (\bm{q_1} +\bm{q_2})  2 i\sin ( \theta_{\bm{q_1}, \bm{q_2}} )     \}
\end{aligned}
\end{equation}
where $\theta_{\bm{q_1}, \bm{q_2}}  = \frac{l^2_B}{2} (q_{1y} q_{2x} -q_{1x} q_{2y} ) $.  
Since  the density operators do not commute   we can estimate the magnitude of the  Trotter error as follows.  Let  
$ ||   \hat{A} ||   \equiv  \text{max}_{ | \Psi \rangle , || | \Psi \rangle  || = 1   }  ||  \hat{A} | \Psi \rangle || $.  
Since the Hamiltonian $\sum_{m } H_m $  is an extensive quantity,   $|| \sum_{m } H_m || \propto N_\Phi$.     Here  $m$  runs over a set of order $N_\Phi$ momenta,  hence implies that   typically,  $ H_m   \propto N_{\Phi}^0 $.  Using this to estimate the  systematic error,   yields the result: 
\begin{equation}
	 f_{QMC}  = f  +  \mathcal{O} \left(   \Delta \tau^2 N_\Phi^2 \right). 
\end{equation}
Hence, to keep the Trotter error under control we have to scale  $\Delta \tau$ as $1/N_\Phi$.

The Trotter error in our model has a  different scaling behavior,  than for models with only local 
interaction   such as the Hubbard model.  For  local interactions  $ || \lambda  || $  scales as  $N_{\Phi} $, 
such that the systematic error on the free energy density is size independent. 


\begin{figure}
\centering
\includegraphics[width=1.0\textwidth]{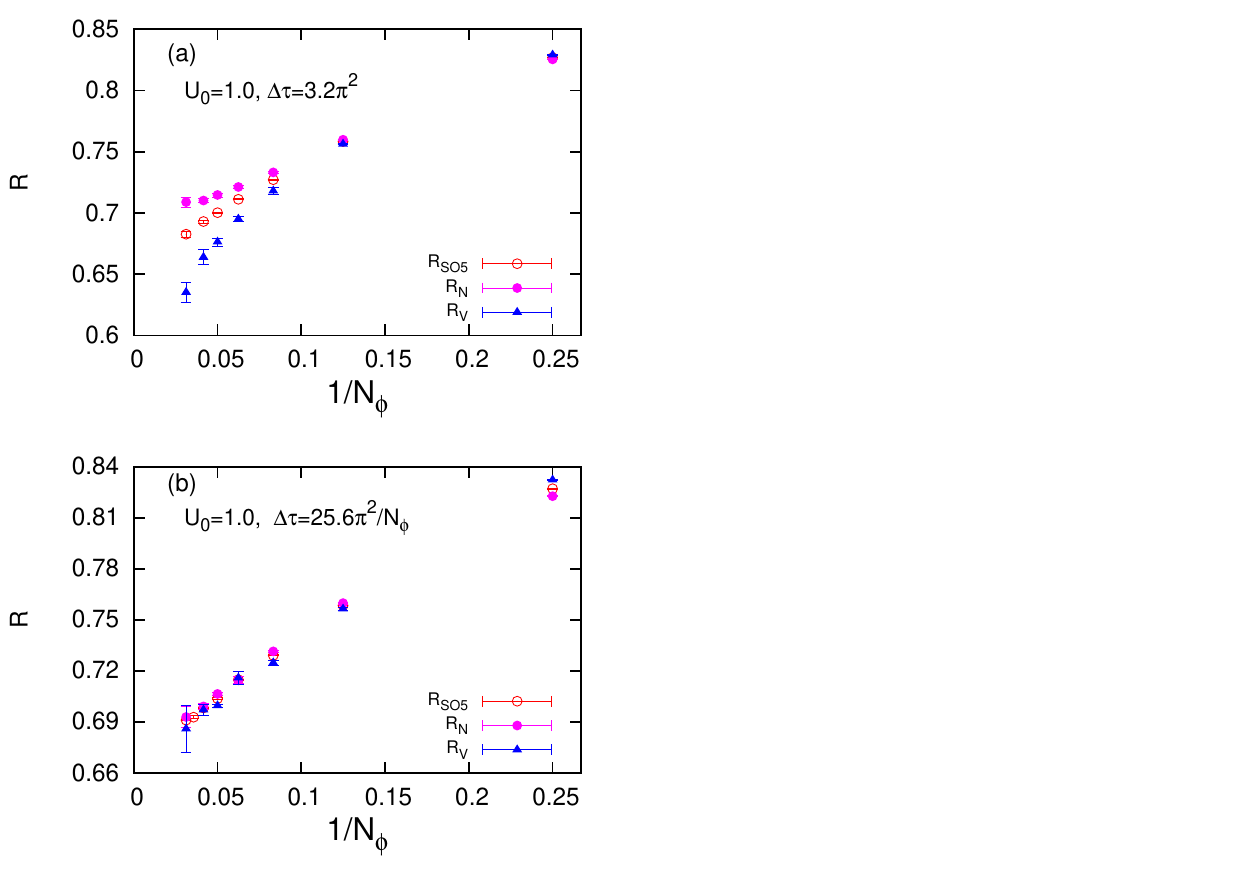}
\caption{
Correlation ratio $R$ of Neel, VBS order and the improved estimator for a fix $\Delta \tau$ as $3.2 \pi^2 $ (a), and for a linear scaling  
of $\Delta \tau = 25.6 \pi^2 /N_{\phi}$ (b). The simulation is based on $U_0=1.0$,   of system sizes $N_{\phi}=4,8,12,16,...32$, with $\beta=1$. }
\label{fig:SO5_Trotter}
\end{figure}

An improved estimator is introduced, based on the SO(5) invariant structure factor: 
\begin{equation}\label{Eq:SO5_Sfactor}
  S(\bm{q})  = \frac{1}{N_{\phi}}  \sum_{i=1}^5  \langle    \hat{n}^i_{\bm{q}} \hat{n}^i_{\bm{-q}}     \rangle.
\end{equation}
The magnetization and correlation ratio used for the scaling analysis in   the main part of the paper is  based on the above structure factor.

Fig.~\ref{fig:SO5_Trotter}  shows a numerical comparison of  the correlation ratio for multiple system sizes. 
In Fig.~\ref{fig:SO5_Trotter} (a)  we consider a  constant  $\Delta \tau$ while in Fig.~\ref{fig:SO5_Trotter} (b)   we scale 
$\Delta \tau$ with the volume:
$\Delta \tau = 25.6 \pi^2 / N_{\phi}$. 
As mentioned previously, our Trotter  decomposition breaks the SO(5) symmetry  such that a convenient measure of the  finite time step systematic error is the discrepancy between the N\'eel and VBS order parameters. 
At  constant $\Delta \tau=3.2 \pi^2$  the correlation ratio defined  from the  N\'eel, VBS  and  SO(5)  order parameters progressively differ as a function of system size. 
On the other hand, for  simulations where we  keep  $\Delta \tau  N_{\phi} $   constant, see Fig.~\ref{fig:SO5_Trotter} (b),  no  $SO(5)$ symmetry 
breaking up to $N_{\phi}=32$ is apparent.   In all our simulations we have kept  $\Delta \tau  N_{\phi}$ constant.

\subsection{Cutoff}

\begin{figure}
\centering
\includegraphics[width=0.45\textwidth]{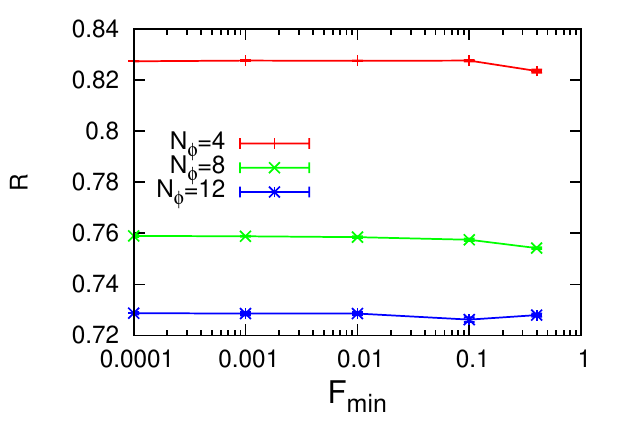}
\caption{ 
Correlation ratio as a function of $F_{min}$ for $N_{\phi}=4,8$ and $12$, at $U_0 =U =1, \beta=160 \pi^2, \Delta {\tau}=3.2 \pi^2$. 
}
\label{fig:U2N2_Cutoff}              
\end{figure}

The effective interacting strength in Eq.~\eqref{Eq:Fierz} is controlled by a momentum dependent function  $F(\bm{q})$ in Eq.~\eqref{Eq:density}:
\begin{equation}
  F(\bm{q}) = e^{ -\frac{1}{4} (q^2_x +q^2_y ) l^2_B }   
\end{equation}  
The exponential  decay  of the interacting strength gives a natural cutoff in the momentum space. In particular, we 
we can consider momenta   satisfying  $F(\bm{q}) > F_{min}$.    
As shown in Fig.~\ref{fig:U2N2_Cutoff}, for $N_{\phi}=4, 8$ and $12$ at $U_0 =U = 1$, the cutoff dependence of   the 
correlation ratio is negligible  up to $F_{min}=0.01$.    In our calculations, we have chosen $F_{min}= 0.01$.  Setting $l_B=1$  implies that the number of $\ve{q}$-vectors  we consider for a given cutoff scales as $N_\phi$.

\subsection{Comparison to exact diagonalisation  }

\begin{figure}
\centering
\includegraphics[width=0.8\textwidth]{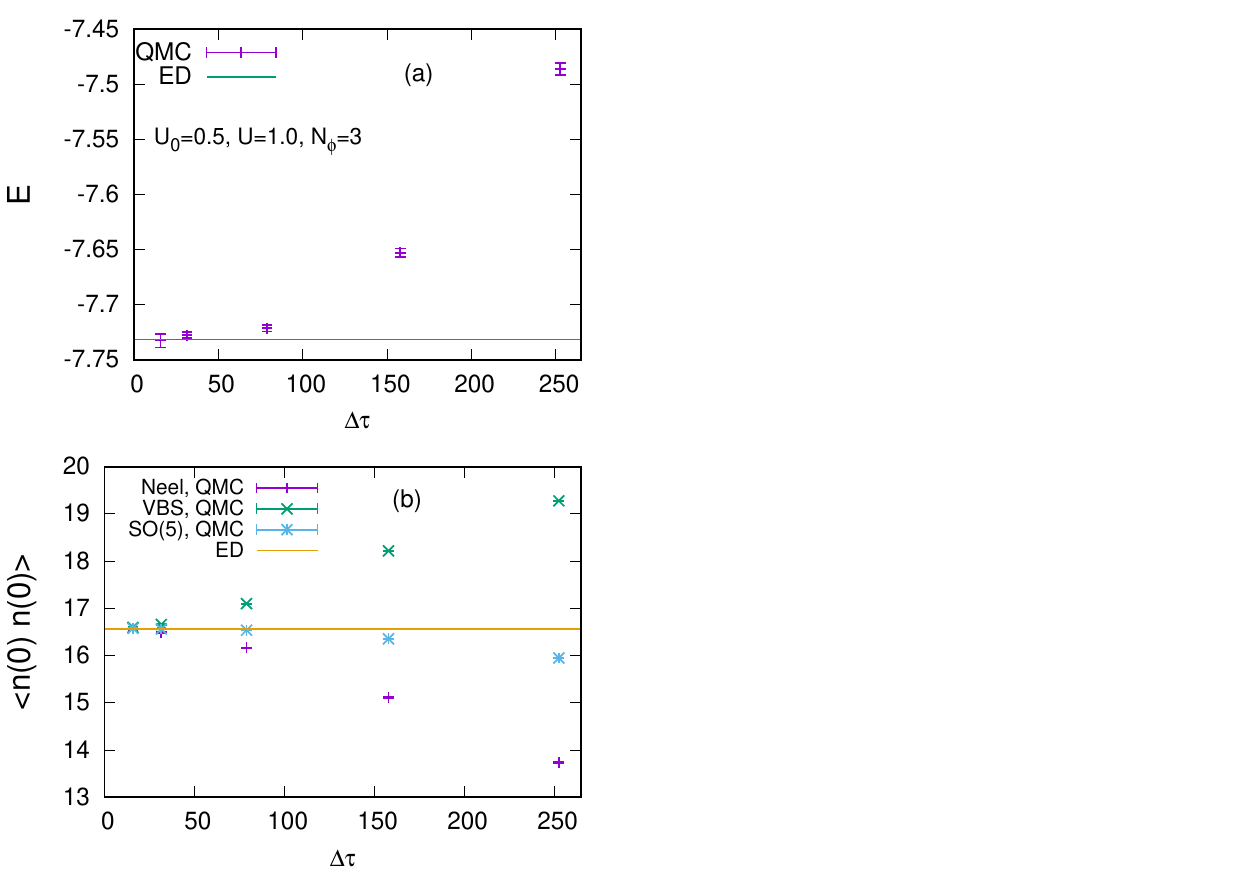}
\caption{
Ground state energy (a) and magnetic order parameter (b) based on AFQMC as a function of $\Delta \tau$, as well as ED. 
The calculation is performed at $U_0=0.5, U=1.0 $ for $N_{\phi}=3$. }
\label{QMC_ED}
\end{figure}

A benchmark calculation of QMC with the exact diagonalization(ED) is performed, based on comparing exact ground state of a 
half filled system, to a  finite temperature  AFQMC simulation at  low enough temperature ($\beta=320 \pi^2 $).
As an example  we consider  $U_0=0.5, U=1.0 $ at $N_{\phi}=3$. 
In Fig.~\ref{QMC_ED},  we see that the two methods show  consistent results for 
the ground state energy and the SO(5) invariant correlation function at zero momentum 
in the limit of small $\Delta \tau$.    
Both the Neel(VBS) correlation function based on the average value of  density operators 
of only the $i=1,2,3$ ($4,5$) term in Eq.~\eqref{Eq:SO5_Sfactor} are equally  shown in Fig.~\ref{QMC_ED}(b).

\subsection{Fitting of the magnetization: proximity to fix-point collision }

Another assumption is that,      the ground state  
in the cases of $U_0 > 0$ is always in a O(5) symmetry breaking phase, with a small magnetization. 
Here we show the fit details based on:. 
\begin{equation}\label{Eq:fit_power_mag}
    m = m_0 + a N_{\phi}^{-\frac{\eta +z}{4} } 
\end{equation}

As we can see from Table~\ref{tab:fit_mag},  the $\chi^2/DOF$ of fit are acceptable, when all the system sizes are included.  
The $\eta$ exponent is robust as  function of $U_0$, except the point of $U_0 =8$.  On the other hand, the extrapolated magnetization become nonzero within  when $U_0 \geq 0.25$.     

\begin{table}
\begin{center}
\begin{tabular}{ |l |l |l |l |l | l| l| }   
\hline
 $ U_0 $    &  $\langle m \rangle_0 $  & $\eta$    &  $\chi ^2$/DOF     \\  \hline
  -1.0      & 0.03(1)                  & 0.33(2)   &   1.51             \\  \hline
  -0.5      & 0.01(1)                  & 0.28(2)   &   1.32             \\  \hline
   0.0      & 0.03(1)                  & 0.29(2)   &   1.78             \\  \hline
   0.25     & 0.028(7)                 & 0.27(1)   &   0.92             \\  \hline
   0.5      & 0.04(1)                  & 0.28(2)   &   0.25             \\  \hline
   1.0      & 0.05(1)                  & 0.28(2)   &   1.17             \\  \hline
   2.0      & 0.064(7)                 & 0.26(2)   &   0.98             \\  \hline
   4.0      & 0.11(1)                  & 0.26(2)   &   1.75             \\  \hline
   8.0      & 0.27(1)                  & 0.38(2)   &   1.11             \\  \hline
\end{tabular}
\end{center}
\caption{\label{tab:fit_mag}
Collective fit  using Eq.~\eqref{Eq:fit_power_mag}. 
}
\end{table}

\end{document}